\documentclass[12pt,preprint]{aastex}

\newcommand{\msun}{$M_{\odot}$}

\def\etal{{et al.\ }}

\begin{document}

\title{Optical Depth of the Cosmic Microwave Background \\
   and Reionization of the Intergalactic Medium }   

\author{J. Michael Shull \& Aparna Venkatesan\altaffilmark{1} }

\affil{University of Colorado, Department of Astrophysical \&
     Planetary Sciences, \\ CASA, 389-UCB, Boulder, CO 80309 }

\altaffiltext{1}{Now at Department of Physics, 2130 Fulton St., 
   University of San Francisco, San Francisco, CA 94117}

\email{mshull@casa.colorado.edu, avenkatesan@usfca.edu}

\begin{abstract}

We examine the constraints on the epoch of reionization 
(redshift $z_r$) set by recent WMAP-3 observations of 
$\tau_e = 0.09 \pm 0.03$, the electron-scattering optical 
depth of the cosmic microwave background (CMB), combined 
with models of high-redshift galaxy and black hole formation.  
Standard interpretation begins with the computed optical depth, 
$\tau_e = 0.042 \pm 0.003$, for a fully ionized medium out to 
$z_r  = 6.1 \pm 0.15$, including ionized helium, which 
recombines at $z \approx 3$.  At $z > z_r$, one must also 
consider scattering off electrons produced by from early black 
holes (X-ray pre-ionization) and from residual electrons left 
from incomplete recombination.  Inaccuracies in computing the 
ionization history, $x_e(z)$ add systematic sources of 
uncertainty in $\tau_e$.  The required scattering at $z > z_r$ 
can be used to constrain the ionizing contributions of ``first 
light" sources.  In high-$z$ galaxies, the star-formation 
efficiency, the rate of ionizing photon production, and the 
photon escape fraction are limited to producing no more than 
$\Delta \tau_e \leq 0.03 \pm 0.03$.  The contribution of 
minihalo star formation and black-hole X-ray preionization at 
$z$ = 10--20 are suppressed by factors of 5--10 compared to 
recent models. Both the CMB optical depth and H~I (Ly$\alpha$) 
absorption in quasar spectra are consistent with an H~I 
reionization epoch at $z_r \approx 6$ providing $\sim50$\% of 
the total $\tau_e$ at $z \leq z_r$, preceded by a partially 
ionized medium at $z \approx$ 6--20.

\end{abstract}

\keywords{cosmology: theory --- cosmic microwave background ---
intergalactic medium }

\newpage

\section*{1. INTRODUCTION }

In recent years, an enormous amount of exciting cosmological data 
have appeared, accompanied by theoretical statements about early 
galaxy formation and the first massive stars.  Many of these 
statements were reactions to first-year (WMAP-1) results (Kogut et al.\
2003; Spergel et al.\ 2003) from the {\it Wilkinson Microwave Anisotropy 
Probe} (WMAP).  These papers inferred a high optical depth to the cosmic 
microwave background (CMB) and suggested early reionization of the intergalactic 
medium (IGM).  Other conclusions came from simplified models and assumptions about 
the stellar initial mass function (IMF), atomic/molecular physics, 
radiative processes, and prescriptions for star formation rates 
and escape of photoionizing radiation from protogalaxies.

In this paper, we focus on the reionization epoch, defined as the 
redshift $z_r$ when the IGM becomes nearly fully ionized over most of 
its volume (Gnedin 2000, 2004).  Our knowledge about reionization comes 
primarily from two types of observations:  H~I (Ly$\alpha$) absorption 
in the IGM and optical depth of the CMB.  
Spectroscopic studies of the ``Gunn-Peterson" (Ly$\alpha$) absorption
toward high-redshift quasars and galaxies imply that H~I reionization
occurred not far beyond $z \sim 6$ (Becker \etal 2001; Fan \etal 2002, 2006),
and that He~II reionization occurred at $z \sim 3$ (Kriss \etal 2001;
Shull \etal 2004; Zheng \etal 2004).  Third-year data from WMAP 
(Spergel \etal 2006; Page \etal 2006) suggest that reionization might 
occur at $z \approx 10$, although large uncertainties remain in 
modeling of the CMB optical depth.  

The WMAP and Ly$\alpha$ absorption results are not necessarily
inconsistent, since
they probe small amounts of ionized and neutral gas, respectively.
In addition, both the H~I absorbers and ionized filaments in
the ``cosmic web" (Cen \& Ostriker 1999) are highly structured 
at redshifts $z < 10$ and affect the optical depths in Ly$\alpha$.  
For example, in order to 
effectively absorb all the Ly$\alpha$ radiation at $z \approx 6$ 
requires a volume-averaged neutral fraction of just 
$x_{\rm HI} \approx 4 \times 10^{-4}$ (Fan \etal 2006).
Simulations of the reionization process (Gnedin 2004; Gnedin \& Fan 2006) 
show that the transition from neutral to ionized is extended in time between 
$z = 5-10$.  The first stage (pre-overlap) involves the development and 
expansion of the first isolated ionizing sources. The second stage marks 
the overlap of the ionization fronts and the disappearance of the last 
vestiges of low-density neutral gas.  Finally, in the post-overlap stage, 
the remaining high-density gas is photoionized.  The final
drop in neutral fraction and increase in photon mean free path occur quite 
rapidly, over $\Delta z \approx 0.3$.

Initial interpretations of the high WMAP-1 optical depth were based on 
models of ``sudden reionization", neglecting scattering from a partially 
ionized IGM at $z > z_r$.  
The revised optical depths, $\tau_e = (0.09-0.10) \pm 0.03$ (Spergel \etal 
2006; Page \etal 2006) also assume a single step to complete ionization 
($x_e = 1$ at redshift $z_r$), although Spergel \etal (2006) explored a 
two-step ionization history, resulting in a broader distribution of $\tau_e$.  
However, the $\chi^2$ curves for two-step ionization provide little 
constraint on the redshift of reionization, or on an IGM with low partial 
ionization fractions, $x_e \ll 0.1$, as we discuss in \S~3. 
Figure 3 of Spergel \etal (2006) shows that the parameters 
$x_e$ and $z_r$ are somewhat degenerate, each with a long tail in the 
likelihood curves.  This emphasizes the possible importance of contributions 
to $\tau_e$ from ionizing UV photons at $z > z_r$ from early massive stars 
and X-rays from accreting black holes (Ricotti \& Ostriker 2004; 
Begelman, Volonteri, \& Rees 2006).

Models of the extended recombination epoch (Seager, Sasselov, \& Scott 2000)
predict a partially ionized medium at high redshifts, owing to residual
electrons left from incomplete recombination.  These electrons produce 
additional scattering, $\Delta \tau_e \approx 0.06$ ($z \approx$ 10--700).  
These effects must be computed in CMB radiation transfer codes such as 
CMBFAST and RECFAST, but only a portion of this scattering affects the large 
angular scales ($\ell \leq 10$) where WMAP detects
a polarization signal.  X-ray preionization can also produce CMB
optical depths $\tau_e \geq 0.01$ (Venkatesan, Giroux, \& Shull 2001,
hereafter VGS01; Ricotti, Ostriker, \& Gnedin 2005).  

In \S~2.1, we show that a fully ionized
IGM from $z$ = 0 to the reionization epoch at $z_r = 6.1 \pm 0.15$ 
(Gnedin \& Fan 2006), produces optical depth, 
$\tau_e \approx 0.042 \pm 0.003$, nearly half the WMAP-3 value.  
Therefore, the high-redshift ionizing sources are limited to 
producing an additional optical depth, $\Delta \tau_e \leq 0.03 \pm 0.03$. 
In \S~3, we discuss the resulting constraints on the amount of star formation 
and X-ray activity at $z \ga 7$ and limits on star formation in minihalos. 
The IGM probably has a complex reionization history, with periods of extended 
reionization for H~I and He~II (Venkatesan, Tumlinson, \& Shull 2003; Cen 2003; 
Wyithe \& Loeb 2003; Hui \& Haiman 2003).  Because $\tau_e$ measures the
integrated column density of electrons, there are many possible scenarios
consistent with the current (WMAP-3) level of CMB data.

\section*{2. OPTICAL DEPTH TO ELECTRON SCATTERING }

\subsection*{2.1. Analytic Calculation of Optical Depth} 

To elucidate the dependence of CMB optical depth on the epoch of reionization 
(redshift $z_r$), we integrate the electron scattering optical depth,
$\tau_e(z_r)$, for a homogeneous, fully ionized medium out to $z_r$.  
For instantaneous, complete ionization at redshift $z_r$, we calculate 
$\tau_e$ as the integral of $n_e \sigma_T d \ell$, the 
electron density times the Thomson cross section along proper length,   
\begin{equation}
   \tau_e(z_r) = \int _{0}^{z_r} n_e \sigma_T (1+z)^{-1} \; [c/H(z)] \; dz \; .   
\end{equation} 
We adopt a standard $\Lambda$CDM cosmology, in which 
$(d \ell /dz) = c(dt/dz) = (1+z)^{-1} [c/H(z)]$, where
$H(z) = H_0 [\Omega_m (1+z)^3 + \Omega_{\Lambda}]^{1/2}$ and
$\Omega_m + \Omega_{\Lambda} = 1$ (no curvature).  
The densities of hydrogen, helium, and electrons are written
$n_H = [(1-Y) \rho_{\rm cr}/m_H](1+z)^3$, $n_{\rm He} = y n_H$,
and $n_e = n_H (1+y)$, if helium is singly ionized.  We also add 
the small ($\tau_e \approx 0.0020$) contribution 
from doubly-ionized helium (He~III) at redshifts $z \leq 3$ (Shull 
\etal 2004). We assume a helium mass fraction by mass $Y = 0.244$ and
define $y = (Y/4)/(1-Y) \approx 0.0807$ (He fraction by number).  The 
critical density is $\rho_{\rm cr} = (1.879 \times 10^{-29}$~g~cm$^{-3}) h^2$
where $h = (H_0$/100~km~s$^{-1}$~Mpc$^{-1}$). 
The above integral can be done analytically:
\begin{equation}
   \tau_e(z_r) = \left( \frac {c}{H_0} \right) \left( \frac {2 \Omega_b}
      {3 \Omega_m} \right)
     \left[ \frac {\rho_{\rm cr} (1-Y)(1+y) \sigma_T } {m_H} \right]
     \left[ \{ \Omega_m (1+z_r)^3 + \Omega_{\Lambda} \}^{1/2} - 1 \right] \; ,  
\end{equation}
where $(1-Y)(1+y) = (1 - 3Y/4)$.  For large redshifts, 
$\Omega_m (1+z)^3 \gg \Omega_{\Lambda}$, and the integral simplifies to
\begin{equation}
  \tau_e(z_r) \approx \left( \frac {c}{H_0} \right) \left( \frac {2 \Omega_b}
   {3 \Omega_m^{1/2}} \right)  \left[ \frac {\rho_{\rm cr} (1-3Y/4) \sigma_T}
   {m_H} \right] (1+z_r)^{3/2} \approx (0.0533) \left[ \frac
    {(1+z_r)}{8} \right] ^{3/2}  \; .   
\end{equation}
Here, we substituted the WMAP-3 parameters 
(Spergel \etal 2006): $\Omega_b h^2 = 0.0223 \pm 0.0009$,
$\Omega_m h^2 = 0.127^{+0.007}_{-0.013}$, and $h = 0.73 \pm 0.03$,
and adopted a primordial helium abundance $Y_p = 0.244 \pm 0.002$
(Izotov \& Thuan 1998; Olive \& Skillman 2001).  A recent paper
(Peimbert \etal 2007) suggests a revised $Y_p = 0.2474 \pm 0.0028$
based on new He~I atomic data.     

From the approximate expression (eq.\ 3), we see that 
$\tau_e \propto (\rho_{\rm cr} \Omega_b \Omega_m^{-1/2} H_0^{-1}$). 
Thus, $\tau_e$ is nearly independent of the Hubble constant, since
$\rho_{\rm cr} \propto h^2$ while the combined parameters, $\Omega_b h^2$ 
and $\Omega_m h^2$, are inferred from D/H, CMB, and galaxy dynamics.  
The scaling with $h$ cancels to lowest order; a slight dependence 
remains from the small $\Omega_{\Lambda}$ term in equation (2).

If we invert the approximate equation (3), we can estimate the primary 
reionization redshift, $(1+z_r) \approx (6.6) [\tau_e(z_r)/0.04]^{2/3}$,
where we scaled to the value, $\tau_e = 0.04$, expected for full
ionization back to $z_r \approx 6$.  As discussed in \S~2.2, this is 
approximately the WMAP-3 value of optical depth ($\tau_e \approx 0.09$) 
reduced by $\Delta \tau_e \approx 0.05$.  This extra scattering may
arise from high-$z$ star formation, X-ray preionization, and 
residual electrons left after incomplete recombination.  The latter 
electrons are computed to have fractional ionization 
$x_e \approx (0.5-3.0) \times 10^{-3}$ between $z$ = 10--700 (Seager \etal 
2000).  Inaccuracies in computing their contribution therefore add
systematic uncertainty to the CMB-derived value of $\tau_e$.  
Partial ionization may also arise from the first stars 
(Venkatesan, Tumlinson, \& Shull 2003, hereafter VTS03) 
and from penetrating X-rays produced by early black holes (VGS01;
Ricotti \& Ostriker 2004, 2005).  These additional 
ionization sources contribute electron scattering that must be subtracted 
from the WMAP-3 values.  They will lower the reionization redshift, 
$z_r = 10.7^{+2.7}_{-2.3}$, derived (Spergel \etal 2006) in the absence 
of partial ionization at $z > z_r$, and they may bring the WMAP-3 
and Gunn-Peterson results into agreement for the epoch of complete
reionization. 

In our calculations, described in \S~3, we make several key assumptions.  
First, we assume a fully ionized IGM out to $z_r \approx 6$, accounting for
both H$^+$ and ionized helium. (Helium contributes 8\% to $\tau_e$, assuming 
He~II at $z > 3$ and an additional $\tau_e \approx 0.002$ for He~III at 
$z \leq 3$).  
Second, we investigate the effects of IGM partial ionization at $z > z_r$.
Finally, in computing the contribution of residual electrons at
high redshifts, we adopt the concordance parameters from the WMAP-3 
data set.  The CMB optical depth is formally only a $3 \sigma$ result,
which may change, as WMAP produces better determinations of the 
matter density, $\Omega_m$, and the parameters, $\sigma_8$ and $n_s$, that 
govern small-scale power.  Both $\sigma_8$ and $n$ have well-known 
degeneracies with $\tau_e$ in CMB parameter extraction 
(Spergel \etal 2006).  Therefore, their derived values may change in
future CMB data analyses, especially as the constraints on $\tau_e$
continue to evolve.  In addition, inaccuracies in the incomplete 
recombination epoch and residual ionization history, $x_e(z)$, add 
uncertainties to the CMB radiative transfer, the damping
of $\ell$-modes, and the polarization signal used to derive an
overall $\tau_e$.

\subsection*{2.2. Residual Electrons in the IGM}

We now discuss the contribution of residual electrons in the IGM following 
the recombination epoch at $z \approx 1000$.  Scattering from these
electrons is significant and is normally accounted for in CMB transport codes 
such as CMBFAST (Seljak \& Zaldarriaga 1996) through the post-recombination
IGM ionization history, $x_e(z)$.  However, a number of past papers are
vague on how the ionization history is treated, which has led to 
confusion in how much residual optical depth and power-damping 
has been subtracted from the CMB signal. 
Modern calculations of how the IGM became neutral have been done 
by Seager \etal (2000), although their code (RECFAST) continues to be modified
to deal with subtle effects of the recombination epoch and the atomic
physics of hydrogen ($2s \rightarrow 1s$) two-photon transitions
(W. Wong \& D. Scott, private communication).  

To illustrate the potential effects of high-$z$ residual electrons,
we have used numbers from Figure 2 of Seager \etal (2000), the 
top-panel model, which assumed a cosmology with 
$\Omega_{\rm tot} = 1$, $\Omega_b = 0.05$, $h = 0.5$, $Y = 0.24$, 
and $T_{\rm CMB} = 2.728$~K.  At low redshifts, $z \approx z_r$, 
just before reionization, they find a residual electron fraction 
$x_{e,0} \approx 10^{-3.3}$.  We fitted their curve for log~$x_e$ 
out to $z \approx 500$ to the formula:
\begin{equation}
  x_e (z) = x_{e,0} \; 10^{0.001(1+z)} \approx 
   (5 \times 10^{-4}) \; \exp [\alpha (1+z)] \; , 
\end{equation}
where $\alpha \approx 2.303 \times 10^{-3}$.  More recent recombination 
calculations (W. Y. Wong \& D. Scott, private communication) using WMAP-3 
parameters ($\Omega_b = 0.04$, $h = 0.73$, $\Omega_m = 0.24$, 
$\Omega_{\Lambda} = 0.76$, $Y = 0.244$) find somewhat lower values, 
$x_{e,0} \approx 10^{-3.67}$ with $\alpha \approx 2.12 \times 10^{-3}$.  
We attribute the lower $x_{e,0}$ to the faster recombination rates arising 
from their higher assumed baryon density, $\Omega_b h^2 = 0.0213$, compared 
to $\Omega_b h^2 = 0.0125$ in Seager \etal (2000).    
 
To compute the electron-scattering of the CMB from these ``frozen-out"
electrons, we use the same integrated optical depth formula (equation 1), 
in the high-$z$ limit, where 
$(d \ell /dz) = (1+z)^{-1} [c/H(z)] \approx (c/H_0) \Omega_m^{-1/2} (1+z)^{-5/2}$.  
We integrate over the residual-electron history, from $z_r \approx 7$ back 
to a final redshift $z_f \gg z_r$, to find
\begin{eqnarray}
 (\Delta \tau_e)_{\rm res} &=& \left( \frac {c} {H_0} \right) 
   \left[ \frac { \rho_{\rm cr} (1-Y) \sigma_T \Omega_b } 
        { \Omega_m^{1/2} m_H } \right]
      \int_{z_r}^{z_f} (1+z)^{1/2} \; x_e(z)  \; dz  \nonumber    \\   
     &=& (3.27 \times 10^{-3}) \; x_{e,o} \int_{(1+z_r)}^{(1+z_f)} 
         u^{1/2} \; \exp (\alpha u) \; du   \; .  
\end{eqnarray}
A rough estimate to the residual scattering comes from setting $\alpha = 0$
and adopting a constant ionized fraction $x_e(z)$, 
\begin{equation}
  (\Delta \tau_e)_{\rm res} \approx (2.36 \times 10^{-3})
     [(1+z_f)^{3/2} - (1+z_r)^{3/2}] \langle x_e \rangle \; .
\end{equation}
This estimate gives $\tau_e = 0.044$ for $z_r = 6$, $z_f = 700$, and 
$\langle x_e \rangle \approx 10^{-3}$.   
More precise values of $\tau_e$ can be derived from the exact integral 
(eq.\ 5) by expanding the exponential as a sum and adopting the limit 
$z_f \gg z_r$,
\begin{equation} 
   (\Delta \tau_e)_{\rm res} = (0.0184) \left[ 
                 \frac {(1+z_f)} {501} \right]^{3/2} \; 
   \sum _{n=0}^{\infty} \frac { [\alpha (1+z_f)]^n }{ (n+3/2) \; n! } \; \; . 
\end{equation}
From $z_r = 7$ out to various final redshifts, we find
cumulative optical depths: 
$(\Delta \tau_e)_{\rm res} = 0.0041$ (from $7 < z < 200$),
$(\Delta \tau_e)_{\rm res} = 0.0088$ (from $7 < z < 300$),
$(\Delta \tau_e)_{\rm res} = 0.0157$ (from $7 < z < 400$),
and $(\Delta \tau_e)_{\rm res} = 0.0256$ (from $7 < z < 500$).     
For $z > 500$, the approximate formula (eq.\ 4) underestimates $x_e$,
but one can integrate the appropriate curves (Seager \etal 2000) 
using piecewise-continuous linear fits.  Between $500 < z < 600$, we 
find $x_e = (2.94 \times 10^{-4}) \exp [0.003224 (1+z)]$, and for
$600 < z < 700$, $x_e = (1.28 \times 10^{-4}) \exp [0.004606 (1+z)]$.   
Integration then yields additional contributions of
$\Delta \tau_e \approx 0.0135$ for $z$ = 500--600 and 
$\Delta \tau_e \approx 0.021$ for $z$ = 600--700.  
These calculations therefore give a total optical depth in residual 
electrons $\tau_e \approx 0.06$ back to $z = 700$.  These electrons
have maximum influence on angular scales with harmonic 
$\ell_{\rm max} \approx 2 z^{1/2} \approx$ 20--50 (Zaldarriaga 1997).  
At higher redshifts, $x_e$ rises to $10^{-2.1}$ at $z = 800$ and to
$10^{-1.1}$ at $z = 1000$, where the CMB source function will affect 
the ``free-streaming" assumption used in CMBFAST (Seljak \& Zaldarriaga 1996).

\section*{3. IMPLICATIONS FOR REIONIZATION MODELS }

The WMAP-3 measurements of fluctuations in temperature ($T$) and
polarization ($E$) have been interpreted to estimate {\it total}
electron-scattering optical depths of $\tau_e = (0.09-0.10) \pm 0.03$.
The central values, $\tau_e = 0.09$ (Spergel \etal 2006) and
$\tau_e = 0.10$ (Page \etal 2006) come, respectively, from computing 
the likelihood function for the six-parameter fit to all WMAP data 
(TT, TE, EE) and for just the EE data as a function of $\tau_e$.
Because of the challenges in translating a single parameter ($\tau_e$)
into a reionization history, $x_e(z)$, it is important to recognize 
the sizable error bars on $\tau_e$. At the 68\% confidence level, 
$\tau_e$ could range from ``low values" (0.06--0.07) up to 
values as high as 0.12--0.13.

In \S~2.1, we showed that $\sim50$\% of this $\tau_e$ can be accounted 
for by a fully ionized IGM at $z \leq z_r$.  
Recent Gunn-Peterson observations of 19 quasars between
$5.7 < z < 6.4$ (Fan \etal 2006; Gnedin \& Fan 2006) are consistent with a
reionization epoch of $z_r = 6.1 \pm 0.15$. According to equation (2),
this produces $\tau_e = 0.042 \pm 0.003$, where our error propagation
includes relative uncertainties in $z_r$ (2.5\%), $\Omega_m h^2$ (8.4\%),
and $\Omega_b h^2$ (4.0\%).  Residual post-recombination electrons
produce a substantial optical depth from $z \approx 10$
back to $z \approx 700$, which uniformly damps all angular scales.
However, their effect on the TE and EE power is considerably
less on large angular scales ($\ell \leq 10$). 
Thus, we can characterize a portion of the 
WMAP-3 observed optical depth, $\tau_e \approx 0.09 \pm 0.03$, through 
known sources of ionization. The ``visible ionized universe"  out to 
$z_r = 6.1$ accounts for $\tau_e = 0.042$, while high-$z$ partial ionization 
could contribute anywhere from $\tau_e \approx 0.01$ to 0.06. 
We therefore assume that an additional optical depth, 
$\Delta \tau_e \leq 0.03 \pm 0.03$, can be attributed to star 
formation and early black hole accretion at $z > z_r$.

Our calculations represent an important change in the derivation of $z_r$ 
from $\tau_e$, suggesting that the amount and efficiency of high-$z$ 
star formation need to be suppressed.  This suggestion is ironic, since WMAP-1
data initially found a high $\tau_e = 0.17 \pm 0.04$ (Spergel \etal 2003)
implying a surprisingly large redshift for early reionization, ranging from 
$11 < z_r < 30$ at 95\% confidence (Kogut \etal 2003).  These results
precipitated many investigations of star formation at $z = 10-30$, some of
which invoked anomalous mass functions, very massive stars (VMS, with $M >
140$~\msun), and an increased ionizing efficiency from zero-metallicity
stars (VTS03; Wyithe \& Loeb 2003; Cen 2003; Ciardi, Ferrara, \& White
2003; Sokasian \etal 2003, 2004). Tumlinson, Venkatesan, \& Shull (TVS04)
disputed the hypothesis that the first stars had to be VMS.  
They showed that an IMF dominated by 10 -- 100~\msun\ stars can produce 
the same ionizing photon budget as VMS, generate CMB optical depths 
of 9--14\%, and still be consistent with nucleosynthetic evidence from
extremely metal-poor halo stars (Umeda \& Nomoto 2003; Tumlinson 2006; 
Venkatesan 2006).

Although the IGM recombination history, $x_e(z)$, is included in calculations 
of CMBFAST and in CMB parameter estimation, the best-fit values of $\tau_e$ 
from WMAP-3 and earlier CMB experiments have been attributed exclusively 
to the contribution from the first stars and/or black holes at $z \leq 20$. 
The contributions from post-recombination electrons ($20 < z < 1100$) have 
not always been subtracted from the data.
This post-recombination contribution was relatively small in some earlier 
models of reionization (Zaldarriaga 1997; Tegmark \& Silk 1995) that explored 
optical depths of $\tau_e = 0.5-1.0$ and suggested reionization epochs up to 
$z_r \sim 100$. 
However, with current data indicating late reionization, it becomes
particularly important to consider contributions to $\tau_e$ prior to the 
first sources of light.

The new WMAP-3 results find a lower $\tau_e$, but they also suggest less
small-scale power available for reionizing sources, owing to lower
normalization parameters, $\sigma_8 \approx 0.74^{+0.05}_{-0.06}$ and
$\Omega_m h^2 \approx 0.127^{+0.007}_{-0.013}$.  This reduction is
somewhat offset by the reduction in spectrum tilt from $n_s = 0.99 \pm
0.04$ (WMAP-1) to $n_s = 0.951^{+0.015}_{-0.019}$ (WMAP-3).  Alvarez
\etal (2006) argue, from the lower values of $\tau_e$ and $\sigma_8$, 
that both WMAP-3 and WMAP-1 data require similar (high) stellar ionizing 
efficiencies.  Haiman \& Holder (2003) use the lower $\tau_e$
to suggest that massive star formation was suppressed in minihalos.   
Our results on a lower $\Delta \tau_e$ make these requirements even 
more stringent, as we now quantify.

Semi-analytic and numerical models of reionization (Ricotti, Gnedin,
\& Shull 2002a,b; VTS03, Haiman \& Holder 2003) show that the efficiency
of ionizing photon injection into the IGM can be parameterized by the
``triple product", $N_{\gamma} f_* f_{\rm esc}$.  Here, $f_*$ represents 
the star-formation efficiency (the fraction of a halo's baryons that go into
stars), $N_{\gamma}$ is the number of ionizing photons produced per baryon
of star formation, and $f_{\rm esc}$ is the fraction of these ionizing
photons that escape from the halo into the IGM.  We can now use our
calculations to constrain the amount of high-$z$ star
formation through the product of these three parameters, henceforth 
referred to as the ``efficiency".  For the ionization history in 
equation (5), we set $x_e(z) = N_{\gamma} f_* f_{\rm esc} c_L(z) f_b(z)$, 
where $c_L(z)$ is the space-averaged baryon clumping factor of ionized
hydrogen, $c_L \equiv \langle n^2_{\rm HII} \rangle
\langle n_{\rm HII} \rangle^2$. We assume
that $c_L$ is the same for H~II and He~III.
The factor $f_b (z)$ is the fraction of baryons in collapsed halos,  
computed through the Press-Schechter formalism (as in VTS03) for the 
cosmological parameters from WMAP-3. We assume that 
$N_{\gamma} f_* f_{\rm esc}$ is constant with redshift.  

There is surely some dependence of each of these parameters on the halo 
mass and environment (Haiman \& Bryan 2006; Ricotti \& Shull 2000). 
Since we have already parameterized the intrahalo recombinations
through $f_{\rm esc}$, we account for the loss of ionizing photons on 
IGM scales through $c_L$ in two forms:  (1) a power-law form with slope 
$\beta = -2$ from the semi-analytic work of Haiman \& Bryan (2006); and
(2) the numerical simulations of Kohler, Gnedin \& Hamilton (2006), using their 
case C (overdensity $\delta \sim 1$ for the large-scale IGM) for $C_R$, 
the recombination clumping factor corresponding to our definition of $c_L$.
In the latter case, the clumping factor is almost constant ($c_L \approx 6$) 
until the very end of reionization. Together, these two different cases 
provide bounds on the range of possible values.

With these assumptions, we can use equation (5) and the allowed 
additional optical depth, $\Delta \tau_e \leq 0.03 \pm 0.03$, 
to constrain the ionizing efficiency of the first stars. In Figure 1, 
we plot the efficiency as a function of $\Delta \tau_e$, for star formation 
in Ly$\alpha$-cooled halos (virial temperature 
$T_{\rm vir} \geq 10^4$ K) and in H$_2$-cooled 
minihalos ($T_{\rm vir} \geq 10^3$ K).  We consider the two clumping factors
from Kohler \etal (2006) and Haiman \& Bryan (2006). 
For $f_\star$ = $f_{\rm esc} = 0.1$, we indicate the efficiencies
corresponding to two cases of interest, assuming a Salpeter initial mass
function (IMF): 
(1) $N_{\gamma} = 60,000$ for a metal-free IMF (10--140 $M_\odot$)
that agrees with both CMB and nucleosynthetic data; and 
(2)  $N_{\gamma} = 4000$ for a present-day IMF (1-100 $M_\odot$). 
Note that $N_{\gamma} = 34,000$ is consistent 
with the WMAP-3 inferrence, $\tau_e \sim 0.1$, if the photons arise from 
zero-metal stars (Tumlinson 2006).  
These values of $N_{\gamma}$ were derived (TVS04) from 
the lifetime-integrated ionizing photon production from various stellar 
populations and IMFs and used as inputs in cosmological reionization models. 

Figure 1 shows that fairly modest efficiencies of massive star formation 
are consistent with limiting the ionizing contribution of minihalos to 
$\Delta \tau_e \leq 0.03-0.06$. This is consistent with suppression  
of star formation and ionizing photon production in mini-halos (Haiman
\& Bryan 2006).   Much larger efficiencies, close to 
those of metal-free stellar populations,  are required for larger halos. 
Interestingly, it seems to make little difference what form is assumed for
the clumping factor, $c_L$, in constraining the ionizing efficiency 
of the first stars.  This is largely a statistical effect arising from
the insensitivity of the {\it average} optical depth, or electron
column density, averaged over many beams passing through a clumpy medium. 

The CMB optical depth can also constrain the level of X-ray preionization 
from high-redshift black holes.  Ricotti \etal (2005) were able to produce 
large optical depths, $\tau_e \approx 0.17$, using accreting high-$z$ black 
holes with substantial soft X-ray fluxes.  Their three simulations 
(labeled M-PIS, M-SN1, M-SN2) produced hydrogen preionization fractions 
$x_e = 0.1-0.6$ between $z = 15$ and $z = 10$, with large co-moving rates 
of star formation, $(0.001-0.1)~M_{\odot}$ Mpc$^{-3}$~yr$^{-1}$, and baryon 
fractions, $\omega_{\rm BH} \approx 10^{-6}$ -- $10^{-5}$ accreted onto black 
holes.  Any significant contribution to $\tau_e$ from X-ray preionization 
requires $x_e \geq 0.1$.  
Therefore, the lower value of $\tau_e$ from WMAP-3 reduces the allowed X-ray 
preionization and black-hole accretion rates significantly  
compared to these models.  Ricotti \etal (2005) demonstrated that high 
optical depths ($\tau_e \approx 0.17$) could be achieved by black-hole X-rays.   
In their models, the IGM at $z > z_r$ was highly ionized ($x_e \gg 0.01$).  
In this limit, 
there were few X-ray secondary electrons and most of the X-ray energy went 
into heating the ionized medium.  By contrast, the WMAP-3 optical depth
suggests that the IGM was much less ionized at $z > z_r$.    

Relating the effects of X-ray ionization from early black holes to a 
$\Delta \tau_e$ and a related X-ray production efficiency factor is 
less straightforward compared to the star formation case, for the 
following reasons.  First, unlike ionization by 
UV photons, X-ray ionization is non-equilibrium in nature and the 
timescales for X-ray photoionization at any epoch prior to $z = 6$
typically exceeds the Hubble time at those epochs (VGS01). Therefore, 
a one-to-one correspondence between X-ray production at an epoch, and 
the average efficiency of halos at that epoch is more difficult to 
establish relative to the UV photon case.  In addition, X-ray 
ionization (whether from stars or black holes), at least initially 
when $x_e <$ 0.1--0.2, will be dominated by secondary ionizations 
from X-ray-ionized helium electrons rather than from direct 
photoionization. This may therefore constrain the physical conditions 
in the IGM (e.g., the level of He ionization) rather than those in the 
parent halo, when we attempt to translate a $\Delta \tau_e$ into an X-ray 
ionization efficiency. Thus, it may be difficult to make precise inferences
about the black hole density and accretion history from $\tau_e$.
 
We define an efficiency parameter, $\epsilon_X$, for X-rays analogous to 
the previous case for massive star formation.  Here, $\epsilon_X$ is the 
product of the average fraction of baryons in black holes (in halos
at $z \ga 7$) and the number of X-ray photons produced (per baryon accreted 
onto such black holes). We assume that the clumping factor ($c_L$)
and escape fraction ($f_{\rm esc}$) are roughly unity for X-rays, given 
their high penetrating power relative to UV photons. 
Thus, we define $x_e(z) = \epsilon_X f_b(z)$ for X-rays, where 
electrons come from H$^+$, He$^+$, and He$^{+2}$.  We assume that each 
X-ray photon produces $\sim$12 hydrogen ionizations, primarily through 
secondary ionizations from X-ray photoelectrons (VGS01).
Figure 2 shows the allowed additional optical depth, 
analogous to the constraints of Figure 1,  
for X-ray efficiency in both Ly$\alpha$-cooled halos and minihalos. 
A comparison with Figure 1 reveals that X-rays are capable 
of much higher ionization efficiency relative to Pop~III or Pop~II 
star formation. 

In summary, we have shown that the revised (WMAP-3) values of
CMB optical depth, $\tau_e = 0.09 \pm 0.03$, lead to a more constrained
picture of early reionization of the IGM.  Approximately half of 
the observed $\tau_e$ comes from a fully ionized IGM back to 
$z_r = 6.1 \pm 0.15$.
The remainder probably arises from the first massive stars and from
accretion onto early black holes at $z > z_r$.  Some of the observed
$\tau_e$ may come from scattering from residual electrons left 
from recombination; inaccuracies in computing this ionization
history add systematic uncertainty to the CMB inferred signal.  
We have assumed extra scattering, $\Delta \tau_e = 0.03 \pm 0.03$
at $z > z_r$, and used this to constrain the effciciencies for
production and escape of ionizing photons, both from the first massive
stars (Figure 1) and early black holes (Figure 2).  

In both cases, the picture is of a partially ionized IGM at redshifts
$z = 6-20$.  For X-ray pre-ionization by early black holes, 
equation (6) can be used to provide an estimate of the effects of partial 
reionization. Between redshifts $z_2 \approx 20$ and $z_1 \approx 6$, 
an IGM with ionized fraction $x_{e,0} = 0.1$  (Ricotti \etal 2005)
would produce $(\Delta \tau_e) = (0.018) (x_{e,0} / 0.1)$,  
which is a significant contribution to the observed $\tau_e = 0.09 \pm 0.03$. 
Ricotti \etal (2005) suggested a large $x_e \approx 0.1-0.6$ and 
pushed their black-hole space densities and accretion rates to large values  
in order to reach the WMAP-1 estimates of $\tau_e = 0.17$.  Because such 
large values of $\tau_e$ are no longer required, the black hole densities 
and IGM ionization fractions are likely to be considerably less.  

All these constraints depend heavily on uncertain parameterizations 
of the efficiency of star formation and ionizing photon production.  
However, with more precise measurements of CMB optical depth
from future missions, there is hope that more stringent constraints on 
high-$z$ star formation and black-hole accretion will be possible.

\section*{Acknowledgements}
We are grateful to David Spergel, Licia Verde, Rachel Bean, and Nick Gnedin 
for useful discussions regarding the interpretation of WMAP data and
numerical simulations.  We thank Douglas Scott and Wan Yan Wong for 
providing their calculations of recombination history.   
This research at the University of Colorado was supported 
by astrophysical theory grants from NASA (NAG5-7262) and NSF (AST02-06042).

\begin{figure}[ht]
  \epsscale{0.9}
  \plotone{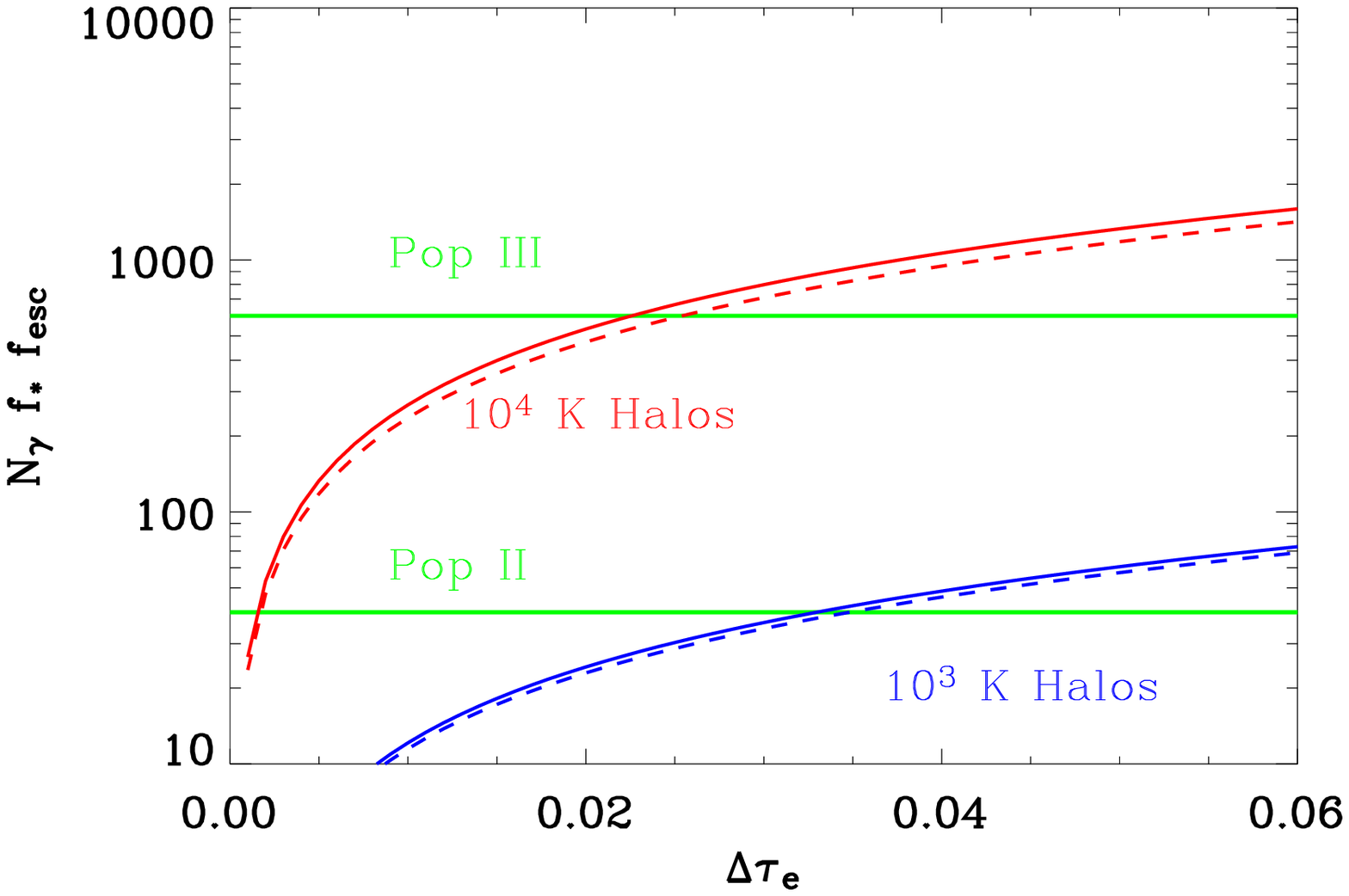}
  \caption{Efficiency factor, $N_{\gamma} f_* f_{\rm esc}$, for production
   and escape of photoionizing radiation vs.\ allowed additional optical
   depth from first stars, $\Delta \tau_e$ at $z > z_r$.  Efficiency
   is defined as in Haiman \& Bryan (2006) for star formation in
   Ly$\alpha$-cooled halos ($T_{\rm vir} \geq 10^4$ K)
   and in H$_2$-cooled minihalos ($T_{\rm vir} \geq 10^3$ K). Solid and
   dashed curves correspond to clumping factors from Kohler \etal (2006) and
   Haiman \& Bryan (2006) respectively. Horizontal solid lines
   correspond to the efficiencies (TVS04) from a metal-free (10--140 $M_\odot$)
   Pop~III IMF and a standard Salpeter (1-100 $M_\odot$) Pop~II IMF,
   with $f_\star$ = $f_{\rm esc}$ = 0.1. See text for discussion. }
\end{figure}

\begin{figure}[ht]
  \epsscale{0.9}
  \plotone{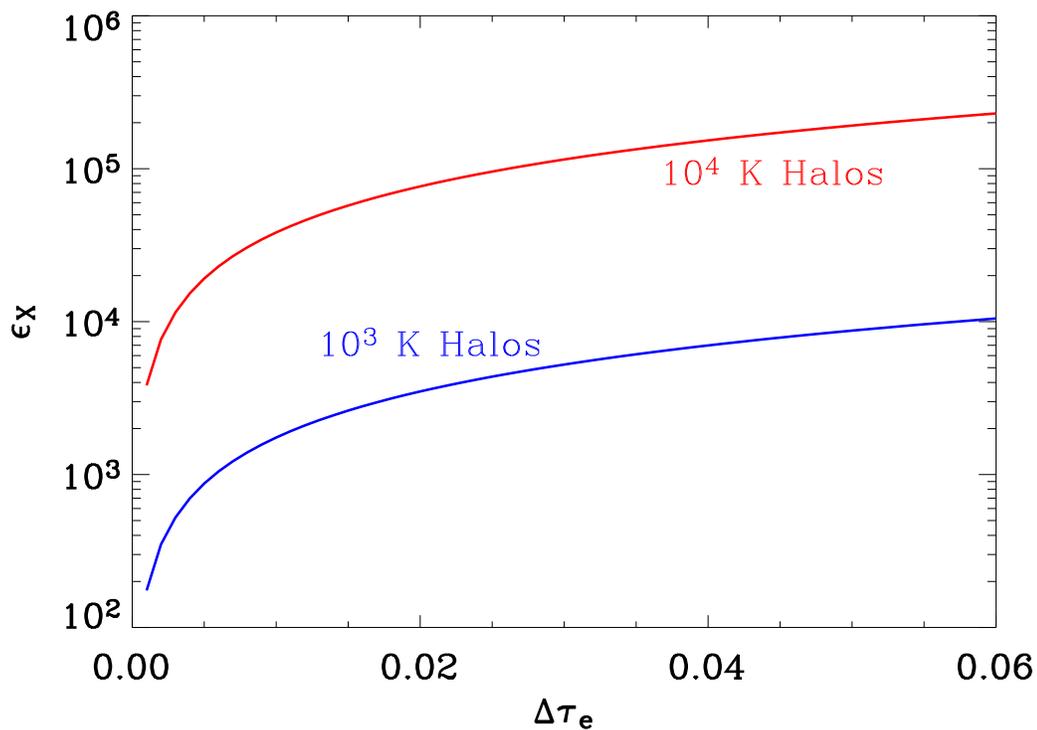}
  \caption{Required X-ray production factor, $\epsilon_X$, 
     versus electron-scattering optical depth, $\Delta \tau_e$, at
     redshifts $z > z_r$.  This factor is taken to be the baryon fraction 
     in black holes times the photons per accreted baryon (see \S~3 for 
     details) and controls the rate of producing escaping X-rays from halos 
     with virial temperatures of $10^3$~K and $10^4$~K.  The electron
     ionized fraction is related to the total baryon fraction by  
     $x_e(z) = \epsilon_X f_b(z)$. } 
\end{figure}

\end{document}